\documentclass[sigconf]{acmart}

\usepackage{booktabs} 
\usepackage{color, colortbl}
\usepackage{booktabs} 
\usepackage{graphicx}
\usepackage{courier}
\usepackage{makecell}
\usepackage{float}
\usepackage{bm}
\usepackage{multirow}
\usepackage[flushleft]{threeparttable}
\usepackage{flushend}
\usepackage{bbm}
\usepackage{mathrsfs}
\usepackage{mwe}
\usepackage{caption}
\usepackage{subcaption}
\usepackage[export]{adjustbox}
\usepackage[ruled, linesnumbered]{algorithm2e} 
\SetKw{KwInput}{Input}
\SetKw{KwOutput}{Output}
\usepackage{bbm}
\usepackage{balance}

\AtBeginDocument{%
  \providecommand\BibTeX{{%
    \normalfont B\kern-0.5em{\scshape i\kern-0.25em b}\kern-0.8em\TeX}}}

\newcommand{\ignore}[1]{}
\newcommand\ChangeRT[1]{\noalign{\hrule height #1}}
\newcommand{\ie}{\emph{i.e., }}

\newcommand{\eg}{\emph{e.g., }}

\usepackage{etoolbox}
\usepackage{tabularx}
\usepackage{enumitem}
\usepackage{multirow}
\usepackage{subcaption}

\setcopyright{acmcopyright}
\copyrightyear{2018}
\acmYear{2018}
\acmDOI{XXXXXXX.XXXXXXX}

\acmConference[Conference acronym 'XX]{Make sure to enter the correct
  conference title from your rights confirmation emai}{June 03--05,
  2018}{Woodstock, NY}
\acmPrice{15.00}
\acmISBN{978-1-4503-XXXX-X/18/06}

\begin{document}


\title{Unconfounded Propensity Estimation for Unbiased Ranking}

%
\author{Dan Luo}
\email{dal417@lehigh.edu}
\affiliation{%
  \institution{Lehigh University}
  \city{Bethlehem}
  \state{PA}
  \country{USA}
}

\author{Lixin Zou$^\dagger$}
\email{zoulixin@whu.edu.cn}
\affiliation{%
  \institution{Wuhan University}
  \city{Wuhan}
  \country{China}
}

\author{Qingyao Ai}
\email{aiqy@tsinghua.edu.cn}
\affiliation{%
  \institution{Tsinghua University}
  \city{Beijing}
  \country{China}
}

\author{Zhiyu Chen$^*$}
\email{zhiyuche@amazon.com}
\affiliation{%
  \institution{Amazon.com, Inc.}
  \city{Seattle}
  \state{WA}
  \country{USA}
}

\author{Chenliang Li}
\email{cllee@whu.edu.cn}
\affiliation{%
  \institution{Wuhan University}
  \city{Wuhan}
  \country{China}
}

\author{Dawei Yin}
\email{yindawei@acm.org}
\affiliation{%
  \institution{Baidu Inc.}
  \city{Beijing}
  \country{China}
}

\author{Brian D. Davison}
\email{davison@cse.lehigh.edu}
\affiliation{%
  \institution{Lehigh University}
  \city{Bethlehem}
  \state{PA}
  \country{USA}
}

\thanks{
$^{\dagger}$ Corresponding author. \\
$^*$ The work was done prior to joining Amazon.
}

\renewcommand{\shortauthors}{Dan Luo et al.}

\begin{abstract}
The goal of unbiased learning to rank~(ULTR) is to leverage implicit user feedback for optimizing learning-to-rank systems. 
Among existing solutions, automatic ULTR algorithms that jointly learn user bias models (\ie propensity models) with unbiased rankers have received a lot of attention due to their superior performance and low deployment cost in practice. 
Despite their theoretical soundness, the effectiveness is usually justified under a weak logging policy, where the ranking model can barely rank documents according to their relevance to the query.
However, when the logging policy is strong, e.g., an industry-deployed ranking policy, the reported effectiveness cannot be reproduced. 
In this paper, we first investigate ULTR from a causal perspective and uncover a negative result: existing ULTR algorithms fail to address the issue of propensity overestimation caused by the query-document relevance confounder.
Then, we propose a new learning objective based on backdoor adjustment and highlight its differences from conventional propensity models, which reveal the prevalence of propensity overestimation.
On top of that, we introduce a novel propensity model called Logging-Policy-aware Propensity (LPP) model and its distinctive two-step optimization strategy, which allows for the joint learning of LPP and ranking models within the automatic ULTR framework, and actualize the unconfounded propensity estimation for ULTR. 
Extensive experiments on two benchmarks demonstrate the effectiveness and generalizability of the proposed method. 
\end{abstract}





\maketitle

\section{Introduction}
\label{section:upe:intro}

Unbiased Learning to Rank (ULTR), leveraging implicit user feedback for optimizing learning-to-rank systems, has been extensively studied in information retrieval~\cite{DBLP:conf/kdd/Joachims02}. 
Usually, directly optimizing ranking models with click data will suffer from the intrinsic noise and bias in user interaction. 
In particular, the position bias~\cite{DBLP:conf/wsdm/CraswellZTR08} that occurs because users are more likely to examine documents at higher ranks is known to distort ranking optimization if not handled properly~\cite{DBLP:conf/sigir/JoachimsGPHG05, DBLP:journals/tois/JoachimsGPHRG07}.
Among existing solutions, ULTR algorithms that jointly estimate click bias and construct unbiased ranking models, namely the AutoULTR algorithms, have drawn a lot of attention~\cite{DBLP:conf/wsdm/WangGBMN18, DBLP:conf/sigir/AiBLGC18, DBLP:conf/www/HuWPL19}. Because they do
not need to conduct separate user studies or online experiments to estimate user bias (\ie the propensity models), and can be deployed on existing systems without
hurting user experiences. 


Despite the theoretical soundness, their effectiveness is usually justified under a weak logging policy, where the ranking model can barely rank documents correctly according to their relevance to the query.  
However, when the logging policy is strong, especially an industry-deployed ranking policy, the reported effectiveness cannot be reproduced. 
Since relevant documents are more likely to be presented on top positions under a strong logging policy, the observed click-through rates on top positions would be greater than those of weak logging policy, \ie users click more on the top positions.
Therefore, the estimated propensity on top positions would be larger than actual values, which we refer to as \textit{propensity overestimation} problem~(Detailed empirical results and analysis in Section~\ref{sec:upe:motivation}).
Since industrial LTR systems will be updated dynamically, implicit feedback will only be valuable when the propensity overestimation is addressed.
%



In this paper, we investigate propensity overestimation in ULTR through the lens of causality. 
By analyzing the causal relations in ULTR, we identify the confounder, \ie query-document relevance, and derive the propensity's compositions in existing ULTR methods: (1) the causal effect between the position and examination, \ie the desired propensity; and (2) the confounding effect caused by the relevance, \ie the overestimated part.
To eliminate the confounding effect, a straightforward solution is to adopt backdoor adjustment~\cite{pearl2000models}, which, in ULTR, means building a propensity model that takes both the position and relevance into account. 
However, optimizing this propensity model is non-trivial because separating ranking and propensity models in AutoULTR algorithms is infeasible when they share a common input~(\ie query-document features) and target~(\ie user clicks)~\cite{DBLP:conf/cikm/YangFLWA20}. 

For unconfounded propensity estimation in ULTR, which we refer to as UPE, we propose a novel propensity model, namely Logging-Policy-aware Propensity Model, and its distinct two-step optimization strategy: 
\textbf{(1) logging-policy-aware confounding effect learning} 
learns a mapping from query-document feature vectors to logging policy scores, which captures the confounding effect caused by the relevance confounder and thereby separates the effects of ranking and propensity model. 
\textbf{(2) joint propensity learning} learns a mapping from the query-document features and position to the examination by locking the confounding effect part and solely optimizing the position-related parameters. 
Thereafter, we are able to conduct unconfounded inference via backdoor adjustment and actualize AutoULTR. 
Extensive experiments on two benchmarks with synthetic clicks on online and offline simulations demonstrate superiority of UPE.

The contributions of this work can be summarized as follows:
\begin{itemize}[leftmargin=*]
    \item We propose the propensity overestimation phenomenon and firstly conduct the 
    causal analysis in ULTR to identify the confounding effect for the overestimation problem.
    \item We propose a novel Logging-Policy-aware Propensity Model and its distinct two-step optimization strategy: (1) logging-policy-aware confounding effect learning and (2) joint propensity learning, which solves the difficulty of backdoor adjustment in ULTR.
    \item We conduct extensive experiments on two benchmarks with synthetic clicks on online and offline simulations to demonstrate the superiority of our proposal.
\end{itemize}
\section{Related Work}
We summarize the related literature on \textit{Unbiased Learning to Rank} and \textit{Deconfounding in Information Retrieval} as follows.

\paragraph{Unbiased Learning to Rank}
To leverage implicit user feedback for optimizing LTR systems, there are two streams for ULTR. 
One school depends on click modeling, which usually maximizes the likelihood of the observed data, models the examination probability and infers accurate relevance feedback from user clicks~\cite{DBLP:conf/wsdm/CraswellZTR08, DBLP:conf/cikm/ChapelleMZG09, DBLP:conf/wsdm/ChuklinMR16, DBLP:MaoLZM18}
However, click models usually requires multiple times appearances of the same query-document pair for reliable inference~\cite{DBLP:conf/ictir/MaoCLZM19}; thus they may fall short for tail queries.
The other school derives from counterfactual learning, which treats bias as a counterfactual factor and debiases user clicks via inverse propensity weighting~\cite{DBLP:conf/wsdm/JoachimsSS17, DBLP:conf/sigir/WangBMN16}. Among them, automatic unbiased learning to rank methods, which jointly learn user bias models~(\ie propensity models) with unbiased rankers, have received a lot of attention due to their superior performance and low deployment cost.


Based on automatic ULTR methods, recent work has investigated various biased within ULTR, including position bias~\cite{DBLP:conf/sigir/AiBGC18, DBLP:conf/cikm/RenTZ22, DBLP:conf/www/HuWPL19, DBLP:conf/kdd/ChenLLS22, DBLP:conf/cikm/VardasbiOR20}, contextual position bias~\cite{DBLP:conf/sigir/ChenLSH21, NEURIPS2022_d81cb1f4, DBLP:conf/www/ZhuangQWBQHC21, DBLP:conf/sigir/YanQZWBN22}, trust bias~\cite{DBLP:conf/cikm/VardasbiOR20, DBLP:conf/cikm/VardasbiRM21}, exploitation bias~\cite{DBLP:conf/sigir/YangLLGYA22}, click unnecessary bias~\cite{DBLP:conf/wsdm/Wang0WW21}, factorizability of the examination hypothesis~\cite{DBLP:conf/kdd/ChenLLS22}. 
Unfortunately, their effectiveness has only been justified under weak logging policies, which cannot be reproduced under a strong policy. 
Our work offers an answer from a causal perspective by identifying relevance as a confounder, and demonstrating a propensity overestimation problem.
Our proposal not only improves the ranking performance of ranking models, but also provides an explanation for the improvement and strong theoretical guarantees of unbiasedness.

\paragraph{Deconfounding in Information Retrieval}
Recently, causal-aware methods have thrived in information retrieval. In particular, some efforts have been made to address confounding problems in recommendation systems. Those methods adopt causal inference to analyze the root causes of bias problems~\cite{DBLP:conf/sigir/ZhangF0WSL021, DBLP:conf/kdd/WangF0WC21, DBLP:conf/cikm/GuptaSMVS21, DBLP:conf/recsys/Christakopoulou20, DBLP:conf/recsys/SatoTSO20} and apply backdoor adjustment~\cite{pearl2000models}  during the training or inference to address the bias problems. For example, \citet{DBLP:conf/kdd/WangF0WC21} identify the distribution of historical interactions as a confounder for bias amplification. \citet{DBLP:conf/sigir/ZhangF0WSL021} identify popularity as a confounder that affects both item exposures and user clicks. However, these methods require the confound variables to be observable, while in ULTR the confounder -- document relevance -- is unobservable. 

There are also a few efforts that address confounding effects without the need to be observable~\cite{DBLP:conf/recsys/WangLCB20, DBLP:conf/recsys/LiuCZDHP021}.  For example, \citet{DBLP:conf/recsys/LiuCZDHP021} learn a biased embedding vector with independent biased and unbiased components in the training phase. In testing, only the unbiased component is used to deliver more accurate recommendations. 
Unlike those methods, extracting separate ranking and propensity models in unbiased learning to rank is difficult when they share a common input and target. 
In summary, these differences make existing deconfounding methods not applicable in ULTR. 

\section{Preliminaries}

In this section, we formulate ULTR with implicit feedback, and introduce the inverse propensity weighting for ULTR.

\subsection{Problem Formulation}

Let $\mathcal{D}$ be the universal set of documents, and $\mathcal{Q}$ be the universal set of queries. 
For a user-issued query $q \in \mathcal{Q}$, $\pi_q$ is the ranked list retrieved for query $q$, $d_k \in \pi_q$ is the document presented at position $k$, $\mathbf{x}_k \in \mathcal{X}$ and $r_k \in \{0, 1 \}$ are the feature vector and binary relevance of the query document pair $(q,d_k)$, respectively.
The goal of learning to rank is to find a mapping function $f$ from a query document feature vector $\mathbf{x}_k$ to its relevance $r_k$. 
In most cases, we are only concerned with the position of relevant documents ($r_d=1$) in retrieval evaluations (\eg MAP, nDCG~\cite{DBLP:journals/tois/JarvelinK02}, ERR~\cite{DBLP:conf/cikm/ChapelleMZG09}), so we can formulate the ideal local ranking loss $\mathcal{L}_{ideal}$ as:  
\begin{equation}
    \label{eq:upe:ideal_loss}
    \mathcal{L}_{ideal}(f,q|\pi_{q}) = \sum_{d_k \in \pi_q, r_k=1}\Delta(f(\mathbf{x}_k),r_k|\pi_{q}),
\end{equation}
where $\Delta$ is a function that computes the individual loss on each document. 
%
An alternative to relevance $r_k$ is implicit feedback from users, such as clicks. 
If we conduct learning to rank by replacing the relevance label $r_k$ with click label $c_k$ in Eq.~\ref{eq:upe:ideal_loss}, then the empirical local ranking loss is derived as follows, 
\begin{eqnarray}
    \label{eq:loss_naive}
    \mathcal{L}_{naive}(f,q|\pi_q) = \sum_{d_k \in \pi_q, c_{k}=1} \Delta(f(\mathbf{x}_k), c_{k}|\pi_q),
\end{eqnarray}
where $c_{k}$ is a binary variable indicating whether the document at position $k$ is cliked in the ranked list $\pi_q$.
However, this naive loss function is biased, due to factors such as position bias~\cite{DBLP:conf/wsdm/JoachimsSS17}.
To address this issue, unbiased learning-to-rank aims to train a ranking model $f$ with the biased user clicks collected but immune to the bias in the implicit feedback.


\subsection{AutoULTR Algorithms}
According to the position bias assumption that the examination only depends on the position, AutoULTR algorithms usually assume the examination and relevance are independent under a weak logging policy. 
Thereafter, according to the \textit{Examination Hypothesis} \cite{DBLP:conf/www/RichardsonDR07}:
\begin{equation}
    c_{k}= 1 \Longleftrightarrow (e_{k}= 1 \ \ \text{and} \ \ r_{k}=1),
\end{equation}
\noindent
the problem of learning a propensity model from click data~(\ie the estimation of bias in clicks) can be treated as a dual problem of constructing an unbiased learning-to-rank model;
and its unbiasedness has been justified in~\citet{DBLP:conf/sigir/AiBGC18}.

Let $e_k$ be the binary variables that represent whether document $d_k$ is examined by a user. In AutoULTR,  an unbiased ranking system $f$ and a propensity model $g$ can be jointly learned by optimizing the local ranking loss as:
\begin{subequations}
    \begin{align}
    \mathcal{L}_{IPW} (f, q|\pi_q) = \sum_{d_k \in \pi_q, c_{k}=1} \frac{\Delta (f(\mathbf{x}_k), c_{k}|\pi_q)}{g(k)} 
    \label{eq:upe:ipw} \\
    \mathcal{L}_{IRW}(g, q| \pi_q) = \sum_{d_k \in \pi_q, c_{k}=1} \frac{\Delta (g(k), c_{k}|\pi_q)}{f(\mathbf{x}_k)}
    \label{eq:upe:irw}
    \end{align}
    \label{eq:upe:causal_cond}
\end{subequations}
where $g(k)$ is the instantiation of $P(e_{k} = 1)$ estimating the propensity at position $k$. 
A nice propensity is that both estimations are only affected by clicked documents $c_k = 1$, respectively.

\section{Causal Analysis on Propensity Overestimation}

\subsection{Causal View of ULTR}
\begin{figure}
     \centering
     \begin{subfigure}[b]{0.20\textwidth}
         \centering
         \includegraphics[width=\textwidth]{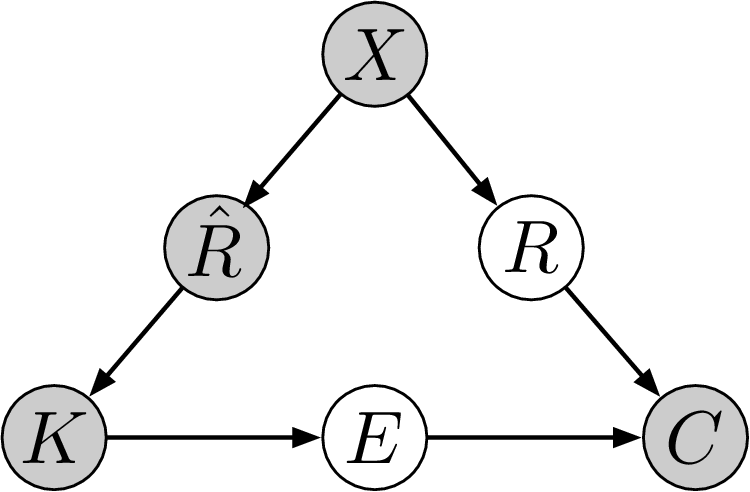}
         \caption{}
         \label{fig:upe:causal_graph}
     \end{subfigure}
     \hspace{1em}
     \begin{subfigure}[b]{0.20\textwidth}
         \centering
         \includegraphics[width=\textwidth]{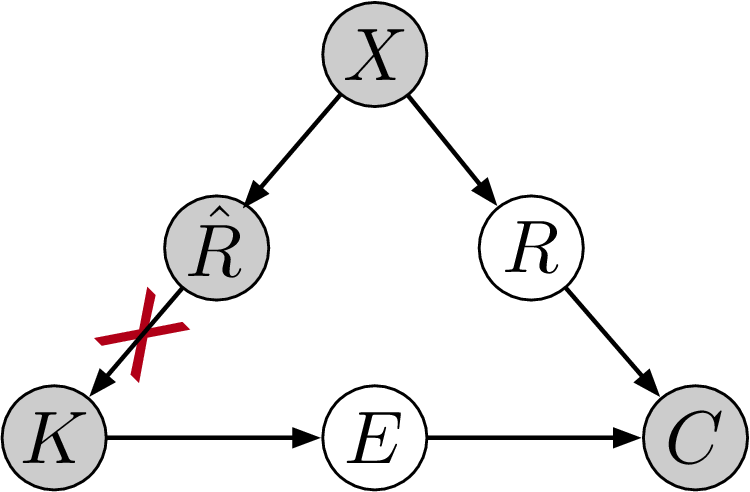}
         \caption{}
         \label{fig:upe:causal_intervention}
     \end{subfigure}
     \vspace{-10pt}
        \caption{(a) The causal graph $G$ for ULTR. (b) The causal graph with intervention $G'$ used in UPE. 
        In particular, we apply backdoor adjustment to cut off the association $X \rightarrow K$ for propensity model learning. A gray node indicates that the variable is observable. 
        }
     \vspace{-15pt}
\end{figure}
To illustrate propensity overestimation in ULTR, we first scrutinize its causal relations and present a causal graph, as shown in Fig.~\ref{fig:upe:causal_graph}, which consists of six variables $\{X, R, \hat{R}, K, E, C \}$ as follows:
\begin{itemize}[leftmargin=*]
    \item $R$ represents the real relevance between the query and document. 
    For simplicity, we assume it is a binary variable with $r \in \{0, 1\}$. However, $R$ is unobservable and it is typically estimated by the ranking model in ULTR algorithms with raw click data.
    \item $X$ is the representation for query-document pair $(q,d)$, where a particular value $\mathbf{x} \in \mathbb{R}^n$ is the feature vector for the pair.
    \item $\hat{R}$ with $\hat{r} \in \mathbb{R}$ is the estimated relevance score generated by the logging policy. 
    \item $K$ denotes $d$'s rank on the search result page, where $K \in \{1..\mathcal{K}\}$
    and $\mathcal{K}$ is the maximum number of observable documents.  
    \item $E$ is a binary variable $e \in \{0, 1 \}$ to indicate whether document $d$ is examined, which is unobservable and typically estimated by the propensity model. 
    \item $C$ is a binary variable $c \in \{0, 1 \}$ to denote whether the user clicked document $d$. 
\end{itemize}

\noindent
The graph's edges describe causal relations between variables:
\begin{itemize}[leftmargin=*]
    \item $X \rightarrow R$: this edge shows that there exists a mapping from query-document representations to their relevance , which is the goal of unbiased learning to rank. 
    \item $X \rightarrow \hat{R}$: feature representation determines the estimated relevance score, based on the logging policy. 
    \item $\hat{R} \rightarrow K$:  the logging policy presents a list of documents in descending order by the estimated relevance scores. Without losing generality, we assume the estimated relevance score of the document decides its ranked position, and ignores the influence of comparing with other documents. 
    \item $K \rightarrow E$: Position bias is formally modeled by assuming the examination only depends on the position.
    \item $(R, E) \rightarrow C$: the edge follows the examination hypothesis~\cite{DBLP:conf/www/RichardsonDR07} that a user would only click a document when it is observed by the user and considered relevant to the user’s need.
\end{itemize}

According to causal theory~\cite{pearl2000models}, query-document feature representation $X$ is a \textit{confounder} as $K \leftarrow X \rightarrow C$, which leads to a spurious correlation in formulating the propensity model.

\subsection{Analysis of Propensity Overestimation}
Based on the causal graph illustrated, we derive the propensity estimand in existing ULTR methods. As discussed previously, they mainly formulate the propensity model conditioned on the clicked items as $P(E|K)$. In causality language, this formulation means: $K$ is the cause, and $E$ is the effect. Thereafter, we can derive propensity estimand $P(E | K)$ as follows:

\begin{subequations}
    \begin{align}
    \underbrace{P(E | K)}_{\text{propensity estimand}}   &= \sum_{\mathbf{x}} P(E, \mathbf{x}|K)   \label{eq:upe:causal_cond_1} \\
        &= \sum_{\mathbf{x}} P(E|\mathbf{x}, K) P(\mathbf{x}|K),    \label{eq:upe:causal_cond_2} \\
    & \varpropto \sum_{\mathbf{x}} \underbrace{P(E|\mathbf{x}, K) P(\mathbf{x})}_{\text{causal}} \cdot \underbrace{P(K|\mathbf{x})}_{\text{confounding}}, \label{eq:upe:compare}
    \end{align}
    \label{eq:upe:causal_cond}
\end{subequations}
where Eq.~\ref{eq:upe:causal_cond_1} is the definition of the law of total probability; Eq.~\ref{eq:upe:causal_cond_2} and Eq.~\ref{eq:upe:compare} follow the Bayes rules, respectively. 
In particular, the proportion operation $\varpropto$ does not affect the effectiveness of the proposal, please refer to Section~\ref{upe:section:pa_finetune} for more details.

We can clearly see the propensity estimand $P(E | K)$ in existing ULTR methods consists of two parts: (1) $P(E|\mathbf{x}, K) P(\mathbf{x})$ that contributes to a \textit{causal effect} between the position and examination, \ie the desired propensity, which will be illustrated in Eq.~\ref{eq:upe:causal_view} later; and (2) $P(K|\mathbf{x})$ that contributes to a \textit{confounding} effect. 
Remarkably, $P(K|\mathbf{x})$ fundamentally changes the propensity estimand $P(E | K)$, especially when users' clicks are collected based on an industry-deployed logging policy. Suppose K is a top position, a more relevant document will have a higher chance to be ranked on position K, resulting in a larger value of $P(K|\mathbf{x})$ for that document. 
Therefore, propensity estimand  $P(E | K)$ on observed data will be larger than the actual value. In short, the query-document relevance confounder $X$ leads to the propensity overestimation.

According to the justification by \citet{DBLP:conf/sigir/AiBGC18}, the unbiasedness of the propensity model in Eq.~\ref{eq:upe:irw} requires the unbiasedness of the ranking model, and vice versa. Since $P(E | K)$ is a biased estimand for the examination, conventional AutoULTR algorithms would fail to converge to the unbiased ranking and propensity models.

Note that this confounding effect widely exists across all logging policies, but its effect is unintentionally concealed in the general ULTR setting, \ie clicks are collected under a weak logging policy, where $P(K|\mathbf{x})$ is almost the same for all documents.
A weak logging policy can barely rank documents correctly according to their relevance to the query, so the confounding effect is minor but non-negligible. 
Experimental results in Section~\ref{section:upe:offline} verify that our solution could obtain a better-performance ranking model even when the backdoor path leads to a minor confounding effect under a weak logging policy.

\section{Methodology}

\begin{figure*}
  \centering
  \includegraphics[width=0.99\linewidth]{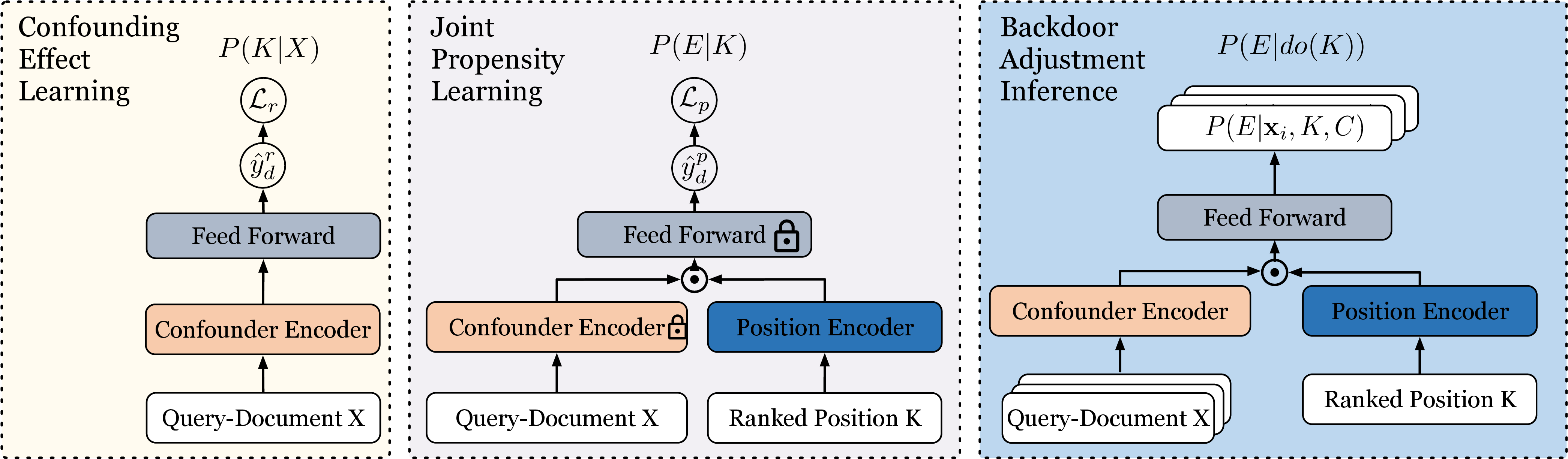}
  \caption{The workflow of the proposed LPP framework. 
    Its optimization strategy consists of two steps: (1) confounding effect learning; (2) joint propensity learning. 
    The unconfounded propensity is obtained by conducting backdoor adjustment over the LPP model.}
  \vspace{-15.pt}
  \label{fig:upe:LPP}
\end{figure*}

\label{section:upe:upe}

In this section, we detail our solution, which is referred to as Unconfounded Propensity Estimation~(UPE).
We first resort to the \textit{backdoor adjustment}~\cite{pearl2000models, DBLP:conf/recsys/WangLCB20}, which enables the causal effect estimation without causal intervention. Then we propose Logging-policy-aware Propensity~(LPP) model, and its novel learning strategy, namely logging-policy-aware confounding effect learning and joint propensity learning. Thereafter, we conduct backdoor adjustment on LPP model for unconfounded propensity inference.

\subsection{Backdoor adjustment}
Experimentally, one can resolve the impact of relevance confounder $X$ by collecting intervened data where the document on specific position is forcibly adjusted to eliminate the impact of the confounder. However, such an experiment is too costly, and faces the risk of doing harm to user experience in practice. Therefore, we resort to the causal technique: \textit{backdoor adjustment}~\cite{pearl2000models}.

According to the theory of backdoor adjustment~\cite{pearl2000models}, our unconfounded propensity estimand is formulated as $P(E | do(K))$, where $do(K)$ can be intuitively seen as cutting off the edge $\hat{R} \rightarrow K$ in $G$, as illustrated in Fig.~\ref{fig:upe:causal_intervention}. 
Remarkably, $P(E | do(K))$ estimates the \textit{causal effect} between position $K$ and examination $E$, rather than $P(E|K)$  estimated by existing methods.  
Now we are ready to derive the specific expression of backdoor adjustment as
\begin{subequations}
    \begin{align}
    P(E | do(K))  &= P_{G'}(E|K)   \label{eq:upe:causal_derive_1} \\
        &= \sum_{\mathbf{x}} P_{G'}(E|\mathbf{x}, K) P_{G'}(\mathbf{x}|K)    \label{eq:upe:causal_derive_3} \\
        &= \sum_{\mathbf{x}} P_{G'}(E|\mathbf{x}, K) P_{G'}(\mathbf{x})      \label{eq:upe:causal_derive_4} \\
        &= \sum_{\mathbf{x}} P(E|\mathbf{x}, K) P(\mathbf{x}),                \label{eq:upe:causal_derive_5}
    \end{align}
    \label{eq:upe:causal_view}
\end{subequations}
\noindent
where $P_{G'}$ denotes the probability function evaluated on intervened causal graph $G'$ in Fig.~\ref{fig:upe:causal_intervention}. In particular, Eq.~\ref{eq:upe:causal_derive_1} is because of \textit{backdoor criterion}~\cite{pearl2000models} as the only backdoor path $K \leftarrow X \rightarrow C$ has been blocked by $do(K)$; 
Eq.~\ref{eq:upe:causal_derive_3} is obtained by Bayes rules; since $K$ and $\mathbf{x}$ are independent in $G'$, $P_{G'}(\mathbf{x}) = P_{G'}(\mathbf{x}|K)$ in Eq.~\ref{eq:upe:causal_derive_4}; in Eq.~\ref{eq:upe:causal_derive_5}, $P(E|\mathbf{x}, K) = P_{G'}(E|\mathbf{x}, K)$ because the causal relation/association $K \rightarrow E$ and $X \rightarrow E$ are not changed when cutting off $\hat{R} \rightarrow K$, and $P(\mathbf{x}) = P_{G'}(\mathbf{x})$ has the same prior on the two graphs. 
To this end, we have demonstrated that the causal component in Eq.~\ref{eq:upe:compare}
exactly contributes to the causal effect, which is the need for unconfounded propensity desired.

Unlike conventional estimand $P(E |K)$, our estimand $P(E | do(K))$ estimates the examination probability for each position with consideration of every possible value of document $\mathbf{x}$ subject to the prior $P(\mathbf{x})$ in Eq.~\ref{eq:upe:causal_derive_5}, rather than $P(\mathbf{\mathbf{x}}|K)$ in Eq.\ref{eq:upe:compare}. 
Therefore, in our method, the documents with high relevance will \textit{not} receive  high examination probability purely because of a higher rank position via $P(\mathbf{x}|K)$, which addresses propensity overestimation.
Remarkably, the ranking model is unbiased at the presence of an unconfounded propensity model and learned by an in-principled unbiased algorithm. 
On the basis of causal theory, our unconfounded estimand learns the causal effect between the position and examination; therefore, its unbiasedness can be guaranteed when jointly works with in-principled unbiased algorithms.

\label{upe:section:back+door_appr}
Theoretically, the sample space of $\mathbf{X}$ is infinite, which makes the calculation of $P(E|do(K))$ in Eq.~\ref{eq:upe:causal_derive_5} intractable. 
Therefore, we further devise an approximation of backdoor adjustment for unconfounded propensity inference by empirically averaging over the training samples as:
\begin{equation}
    \label{eq:upe:do_cal}
    P(E|do(K)) = \frac{1}{|\mathcal{Q}_b| \cdot |\pi_q|}\sum_{q \in \mathcal{Q}_b} \sum_{d_i \in \pi_q} P(E|\mathbf{x}_i, K),
\end{equation}
\noindent where $\mathcal{Q}_b$ is a batch of queries in the raw click data, $\pi_q$ is the ranked list for query $q$, $|\mathcal{Q}_b|$ and $|\pi_q|$ denotes the number of queries within the batch and number of documents in the ranked list, respectively.

\subsection{Logging-Policy-aware Propensity Model}
To facilitate the backdoor adjustment in Eq.~\ref{eq:upe:do_cal}, we need to instantiate a propensity model as $P(E |X, K)$. However, it is difficult to extract separate ranking models and propensity models when they share the common input~(\ie document) and target~(\ie user clicks)~\cite{DBLP:conf/cikm/YangFLWA20}. 
Inspired by the derivation in Eq.~\ref{eq:upe:causal_cond}, we propose a novel propensity model that considers both the relevance and position, named Logging-Policy-aware Propensity Model~(LPP) (as shown in Fig.~\ref{fig:upe:LPP}), and its novel two-step optimization strategy: logging-policy-aware confounding effect learning and joint propensity learning.



Our model consists of three parts: a confounder encoder, a position encoder and shared feed-forward network. 
The encoders learn a dense representations for each query-document pair and its ranking position, respectively. 
Afterward, a vector sum integrates relevance and position representations. The feed-forward network receives the output from the encoder(s) and generates the final output for optimization with respect to different targets.
In particular, we discuss the design choice of LPP model as follows.  

\textbf{Observable confounder variable}.
Existing deconfounding methods~\cite{DBLP:conf/sigir/ZhangF0WSL021, DBLP:conf/kdd/WangF0WC21, DBLP:conf/sigir/YangFJWC21, DBLP:conf/kdd/ZhanPSWWMZJG22} usually represent the joint effect by a simple multiplication among each decoupled factor. Unfortunately, it is not feasible in ULTR because neither $P(E|K)$ nor $P(E|\mathbf{x})$ is observable. 
In LPP, we leverage a shared feed-forward network to capture the joint effect of relevance features $X$ and position $K$. 

\textbf{The motivation behind LPP model}.
At first glance, readers may confuse the LPP in our work and contextual position bias~\cite{DBLP:conf/sigir/YanQZWBN22, DBLP:conf/www/ZhuangQWBQHC21, DBLP:conf/sigir/ChenLSH21}, since they both model $P(E|X, K)$. 
The key difference between them is: the propensity model desired in our work is $P(E|do(K))$ according to the position-based bias assumption; whereas it is $P(E|X, K)$ according to the contextual position bias. 
In this work, we need to build a propensity model $P(E|X, K)$  because we need to address propensity overestimation caused by the relevance confounder as illustrated in Eq.~\ref{eq:upe:causal_view}. More importantly, we need to separate the influence over propensity estimation by the position and relevance from raw click data, which facilitates backdoor adjustment and obtain the unconfounded propensity estimation $P(E|do(K))$.



\subsubsection{Logging-policy-aware Confounding Effect Learning}
\label{section:upe:dicussion}
The first step is to estimate the confounding effect caused by query-document features $X$ under a logging policy, which corresponds to $P(K|X)$ in Eq.~\ref{eq:upe:compare}. To this end, we propose logging-policy-aware confounding effect learning, which learns a mapping from raw document features to logging policy scores $\hat{R}$. Careful readers may notice that the target is logging policy scores $\hat{R}$ instead of rank positions $K$. We will compare the effectiveness of different fitting targets in Section~\ref{section:upe:target}.

Directly learning the mapping in a point-wise way, however, is problematic. Since the logging policies, \ie ranking models, are usually optimized pairwise or listwise, the logging policy scores may follow different distributions under different queries. This issue would restrict the expressive ability of neural encoders.

Therefore, we propose to learn the mapping in a list-wise way, which is invariant to different score distributions under different queries. Formally, given a query $q$ and its associated documents $\pi_q = [d_1, \cdots, d_N]$, each feature vector for query-document pair $\mathbf{x}_i$ is transformed into an $n$-dimensional latent feature representation:
\begin{equation}
    \mathbf{m}_{d_i} = \text{Encoder}_D (q, d_i), \ \ \text{where } \mathbf{m}_{d_i} \in \mathbb{R}^n.
\end{equation}
\noindent
$\text{Encoder}_D$ is a confounder encoder that projects the raw query-document feature vector $\mathbf{x}_i$ to latent representations $\mathbf{m}_{d_i}$.

Afterward, the latent feature vector is passed into a feed-forward network, and generate the predictive scores as:
\begin{equation}
    \label{eq:upe:ffn}
    \hat{y}^r_{d_i} = \text{FFN}(\mathbf{m}_{d_i}), \ \ \text{where }  \hat{y}^r_{d_i} \in \mathbb{R},
\end{equation}
$\text{FFN}$ is a point-wise feed-forward network that projects the latent representation to real-value scalar as prediction scores $P(K|\mathbf{x}_i)$. 

Let $y^r_{d_i}$ be the logging policy score of document $d_i$ in query $q$ that is observed in the logging policy, we optimize in the list-wise paradigm through an attention rank loss function~\cite{DBLP:conf/sigir/AiBGC18}:
\begin{equation}
    \label{eq:upe:pretrain_loss}
    \mathcal{L}_r(g_{\text{pt}} | \pi_q) = -\sum_{d_i \in \pi_q} \frac{\exp (y^r_{d_i})}{\sum_{z \in \pi_q}{\exp(y^r_z)}} \cdot \log \frac{\exp (\hat{y}^r_{d_i})}{\sum_{z \in \pi_q}{\exp(\hat{y}^r_z)}},
\end{equation}
where $g_{\text{pt}}$ denotes the parameters including in the confounder encoder $\text{Encoder}_D$ and feed-forward network FFN. 
To sum up, the logging-policy-aware confounding effect learning captures confounding component $P(K|X)$ in Eq.~\ref{eq:upe:compare}, and enables the separation of the ranking and propensity model from raw clicks.


\textbf{A powerful confounder encoder is needed}.
As shown in Eq.~\ref{eq:upe:compare}, a more accurate estimation of the confounding effect naturally leads to a more better unconfounded propensity models, given a logging policy. 
Thus, an expressive confounder encoder is needed for separating the ranking and propensity models from raw clicks, and consequently, the performance of ranking models can be enhanced.

\subsubsection{Joint Propensity Learning}
\label{upe:section:pa_finetune}

Next, we present how to capture the influence of position $K$.
In particular, we propose to learn the mapping from position and query-document feature to the examination by fixing the confounding effect model and solely
tuning the position-related parameters, as marked lock in Fig.~\ref{fig:upe:LPP}.
The rationality is that fixing the confounder encoder and feed-forward network keeps the relevance confounding effect unchanged; therefore, the position encoder is able to correctly capture the influence of position over examination. 

Formally, we design a position embedding function  $\text{Encoder}_P$. It encodes the position of a document that is ranked on position $k$ to a vector with the same dimension as $\mathbf{m}_{d_i}$: 
\begin{equation}
    \mathbf{p}_{k} = \text{Encoder}_P(\text{rank}(d_k)), \ \ \text{where } \mathbf{p}_{k} \in \mathbb{R}^n,
\end{equation}
and $\text{rank}(d_k)$ is the ranked position for document $d_k$ generated by the logging policy. 
Afterward, we obtain predictions for joint propensity scores $P(E | K)$ through the \textit{frozen} $\text{FFN}$: 
\begin{equation}
    \hat{y}^p_{k} = \text{FFN}(\mathbf{m}_{d_k} + \mathbf{p}_{k}).
\end{equation}

Let $y^p_{k}$ be the estimated propensity score of the document ranked at position $k$ by existing ULTR algorithms, which only depends on position $k$. We solely update the position embedding function $\text{Encoder}_P$ during the optimization
via the attention rank loss function, which is formally defined as,
\begin{equation}
    \label{eq:upe:tuning_loss}
        \mathcal{L}_p (g_{\text{pos}} | \pi_q) = -\sum_{d_k \in \pi_q} \frac{\exp (y^p_k)}{\sum_{z \in \pi_q}{\exp(y^p_z)}} \cdot \log \frac{\exp (\hat{y}^p_k)}{\sum_{z \in \pi_q}{\exp(\hat{y}^p_z)}},
\end{equation}
where $g_{\text{pos}}$ denotes the parameters of position embedding $\text{Encoder}_P$. To this end, the joint propensity learning captures the influence of position $K$ by fixing that of confounder $X$.

As shown in Eq.~\ref{eq:upe:pretrain_loss} and Eq.~\ref{eq:upe:tuning_loss}, the use of the softmax function assumes that the relevance probabilities and examination probabilities on different documents in $\pi_q$ will sum up to 1, which is not true in practice. This, however, does not hurt the effectiveness of models training. In fact, the predicted values of $\hat{y}^p_k$ have a minor effect on the unconfounded propensity learning as long as their relative proportions are correct. Due to this reason, we show normalized propensity against position 10, which reflects the relative proportion, throughout this paper.
Such technique has been widely applied in existing work, and its effectiveness has been extensively verified in prior work~\cite{DBLP:conf/sigir/AiBLGC18, DBLP:conf/cikm/CaiGFAZC22, DBLP:journals/corr/abs-2207-11785}. 

\begin{algorithm}
\caption{Unconfounded Propensity Estimation for Unbiased Learning to Rank}\label{alg:upe}
\SetKwInOut{Input}{Input}
\SetKwInOut{Output}{Output}

\Input{query set $\mathcal{Q}$, an ULTR algorithm $F$}
\Output{ranking model $f$ and  logging-policy-aware propensity model $g$}

\For{number of steps for \textbf{training the unbiased ranking model}} {
    Sample a batch of queries $Q$ from $\mathcal{Q}$ with ranked lists $\pi_q$ and logging policy scores $y^r$ per query $q$ \;
    Estimate propensity scores $y_k^p$ based on existing ULTR algorithm $F$\;
    Estimate confounding effect $P(K|X)$ by updating the confounder encoder and feed-forward network in LPP model according to Eq.~\ref{eq:upe:pretrain_loss} \;
    Freeze the parameters in the confounder encoder and feed-forward network in LPP model\;
    Position locked-tuning by updating the position encoder according to Eq.~\ref{eq:upe:tuning_loss}\;
    Infer unconfounded propensity $P(E|do(K))$ according to backdoor adjustment approximation according to Eq.~\ref{eq:upe:do_cal}\;
    Based on unconfounded propensity $P(E|do(K))$, update ranking model $f$, according to Eq.~\ref{eq:upe:ipw}.
}
\end{algorithm}

\subsubsection{Unconfounded Propensity Inference with Backdoor Adjustment}
Given the LPP model, we first estimate the propensity under query-document relevance confounder:
\begin{equation}
    \begin{aligned} 
        P(E|\mathbf{x}_i, K) := \text{FFN} \big( \text{Encoder}_D(\mathbf{x}_i) + \text{Encoder}_P(k) \big).
    \end{aligned}
\end{equation}
\noindent 
Afterward, we leverage backdoor adjustment approximation in Eq.~\ref{eq:upe:do_cal} to obtain unconfounded propensity estimation $P(E|do(K), C)$ from raw user clicks, and integrate with any IPW-based ULTR algorithm to obtain unbiased ranking models. 
Given the uncoufounded propensity score, one can conduct any IPW-based ULTR algorithm with our proposal, as it is a plug-in model, which can seamlessly integrated into existing automatic ULTR framework. We summarize UPE for ULTR in Algorithm~\ref{alg:upe}. 

\section{Experiments}
We conduct extensive experiments to demonstrate the effectiveness of our UPE by investigating the following research questions:
\begin{itemize}[leftmargin=*]
    \item \textbf{RQ1}: Can empirical results justify the propensity overestimation problem and how it affects ranking models?
    \item \textbf{RQ2}: Can UPE address the propensity overestimation problem and obtain an unbiased ranking model with dynamic training and serving of LTR systems?
    \item \textbf{RQ3}: Can UPE still address the propensity overestimation problem and obtain an unbiased ranking model on classic ULTR settings, \ie the offline learning paradigm?
    \item \textbf{RQ4}: What influence do variant designs have on optimizing the logging-policy-aware propensity model, and the ranking model?
\end{itemize}
In particular, we analyze two learning paradigms that are proposed  in~\citet{DBLP:journals/tois/AiYWM21}.  The first one, referred to as the deterministic online paradigm ($\mathit{OnD}$), is an online setting where the displayed ranked list $\pi_q$ is created by the current logging policy, and the ranking model is updated based on $c_{\pi_q}$ collected online. 
The second one, which is referred to as the offline paradigm ($\mathit{Off}$), is a classic setting where we obtain a logging policy, and then both the displayed ranked list $\pi_q$ and the clicks on it $c_{\pi_q}$ are fixed and observed in advance. 
{\centering
\tabcolsep 0.02in
\begin{table*}[!hpt]
\centering
\caption{Overall performance comparison between UPE and the baselines on Yahoo! and Istella-S datasets with deterministic online learning~($\mathit{OnD}$). ``$\ast$''  indicates statistically significant improvement over the best baseline without result randomization.}
\vspace{-10pt}
\label{tab:upe:online_exp}
    \centering
    \begin{tabular}{ c c c c c  c c c c c  c c c c c  c c}
    \ChangeRT{0.8pt}
    \multirow{3}{*}{\shortstack{Methods}} & \multicolumn{8}{c}{ Yahoo! LETOR } & \multicolumn{8}{c}{ Istella-S } 
    \\ \cmidrule(lr){2-9} \cmidrule(lr){10-17}
    & \multicolumn{4}{c}{ NDCG@K } & \multicolumn{4}{c}{ ERR@K } & \multicolumn{4}{c}{ NDCG@K } & \multicolumn{4}{c}{ ERR@K }  \\
    \cmidrule(lr){2-5}  
    \cmidrule(lr){6-9}      
    \cmidrule(lr){10-13}  
    \cmidrule(lr){14-17} 
    & K = 1 & K = 3 & K = 5 & K = 10 & K = 1 & K = 3 & K = 5 & K = 10  & K = 1 & K = 3 & K = 5 & K = 10 & K = 1 & K = 3 & K = 5 & K = 10 \\
    \ChangeRT{0.5pt} 
    IPW-Random    & 0.693 & 0.702 & 0.722 & 0.767 & 0.352 & 0.432 & 0.452 & 0.469 
           & 0.667 & 0.637 & 0.660 & 0.720 & 0.596 & 0.705 & 0.722 & 0.728 \\
    \ChangeRT{0.5pt}
    UPE    & $\textbf{0.695}^\ast$ & $\textbf{0.703}^\ast$ & $\textbf{0.723}^\ast$ & $\textbf{0.768}^\ast$
           & $\textbf{0.355}^\ast$ & $\textbf{0.433}^\ast$ & $\textbf{0.454}^\ast$ & $\textbf{0.469}^\ast$ 
           & $\textbf{0.666}^\ast$ & $\textbf{0.633}^\ast$ & $\textbf{0.657}^\ast$ & $\textbf{0.718}^\ast$
           & $\textbf{0.595}^\ast$ & $\textbf{0.704}^\ast$ & $\textbf{0.720}^\ast$ & $\textbf{0.727}^\ast$  \\ 
    Vectorization & 0.670 & 0.678 & 0.702 & 0.753 & 0.343 & 0.423 & 0.446 & 0.460
                  & 0.663 & 0.630 & 0.653 & 0.711 & 0.593 & 0.701 & 0.716 & 0.724 \\
    DLA    & 0.671 & 0.677 & 0.701 & 0.750 & 0.345 & 0.423 & 0.445 & 0.460 
           & 0.663 & 0.629 & 0.653 & 0.712 & 0.592 & 0.701 & 0.717 & 0.724 \\
    REM    & 0.674 & 0.678 & 0.699 & 0.747 & 0.349 & 0.425 & 0.446 & 0.462 
           & 0.642 & 0.611 & 0.631 & 0.690 & 0.574 & 0.684 & 0.702 & 0.709 \\
    PairD  & 0.602 & 0.614 & 0.642 & 0.700 & 0.319 & 0.394 & 0.416 & 0.433 
           & 0.609 & 0.569 & 0.593 & 0.653 & 0.545 & 0.656 & 0.675 & 0.684 \\
    Naive  & 0.634 & 0.644 & 0.670 & 0.723 & 0.334 & 0.409 & 0.431 & 0.447 
           & 0.639 & 0.601 & 0.624 & 0.683 & 0.571 & 0.681 & 0.699 & 0.706 \\
    \ChangeRT{0.8pt}
    \end{tabular}
\end{table*}}

\subsection{Experimental Settings}
\paragraph{Datasets}
We conduct empirical studies on two of the largest publicly available LTR datasets: 
\begin{itemize}[leftmargin=*]
    \item \textbf{Yahoo! LETOR}\footnote{\url{https://webscope.sandbox.yahoo.com/}} comes from the Learn to Rank Challenge version 2.0 (Set 1), and is one of the largest benchmarks of unbiased learning to rank~\cite{DBLP:conf/sigir/AiBLGC18, DBLP:conf/www/HuWPL19}. It consists of 29,921 queries and 710K documents. Each query-document pair is represented by a 700-D feature vector and annotated with 5-level relevance labels~\cite{DBLP:journals/jmlr/ChapelleC11}.
    \item \textbf{Istella-S}\footnote{\url{http://quickrank.isti.cnr.it/istella-dataset/}} contains 33K queries and 3,408K documents (roughly 103 documents per query) sampled from a commercial Italian search engine. Each query-document pair is represented by 220 features and annotated with 5-level relevance judgments~\cite{DBLP:conf/sigir/LuccheseNOPST16}.
\end{itemize}
We follow the predefined data split of training, validation, and testing of all datasets. The Yahoo!\ set splits the queries arbitrarily and uses 19,944 for training, 2,994 for validation, and 6,983 for testing. The Istella-S dataset has been divided into train, validation, and test sets according to a $60\% - 20\% - 20\%$ scheme.

\paragraph{Click Simulation}
We generate synthesized click with a two-step process as in \citet{DBLP:conf/wsdm/JoachimsSS17} and \citet{DBLP:conf/sigir/AiBLGC18}. First, we generate the initial ranked list $\pi_q$ for each query $q$ based on learning paradigms, \ie $\mathit{OnD}$ and $\mathit{Off}$. 
Then, we simulate the user browsing process based on PBM~\cite{DBLP:conf/wsdm/JoachimsSS17} and sample clicks from the initial ranked list by utilizing the simulation model. 
The PBM models user browsing behavior based on the assumption that the bias of a document only depends on its position: 
\begin{equation}
    \label{eq:upe:rho}
    P(e_{d_i}) = \rho_i^\eta, 
\end{equation}
where $\rho_i$ represents position bias at position $i$ and $\eta \in [0, +\infty]$ is a parameter controlling the degree of position bias. The position bias $\rho_i$
is obtained from an eye-tracking experiment in~\citet{DBLP:conf/wsdm/JoachimsSS17} and the parameter $\eta$ is set as 1 by default. Following the methodology proposed by ~\citet{DBLP:conf/cikm/ChapelleMZG09}, we sample clicks with: 
\begin{equation}
    \Pr (r_{d_i} = 1 | \pi_q) = \epsilon + (1 - \epsilon) \frac{2 ^ y - 1}{2^{y_{\mathrm{max}}} - 1},
\end{equation}
where $y \in [0, y_{\mathrm{max}}]$ is the relevance label of the document $d_i$, and $y_{\mathrm{max}}$ is the maximum value of $y$, which is 4 on both datasets. $\epsilon$~is the noise level, which models click noise such that irrelevant documents (\ie $y = 0$) have non-zero probability to be perceived as relevant and clicked. We fix $\epsilon = 0.1$ as the default setting.

\paragraph{Baselines}
To demonstrate the effectiveness of our proposed method, we compare it with baseline methods which are widely used in ULTR problems.
\begin{itemize}[leftmargin=*]
  \item \textbf{Vectorization}: The Vectorization~\cite{DBLP:conf/kdd/ChenLLS22} expands the examination hypothesis to a vector-based one, which formulates the click probability as a dot product of two vector functions instead of two scalar functions.
  \item \textbf{DLA}: The Dual Learning Algorithm~\cite{DBLP:conf/sigir/AiBLGC18} treats the problem of unbiased learning to rank and unbiased propensity estimation as a dual problem, such that they can be optimized simultaneously.  
  \item \textbf{REM}: The Regression EM model~\cite{DBLP:conf/wsdm/WangGBMN18} uses an EM framework to estimate the propensity scores and ranking scores. 
  \item \textbf{PairD}: The Pairwise Debiasing (PairD) Model~\cite{DBLP:conf/www/HuWPL19} uses inverse propensity weighting for pairwise learning to rank. 
  \item \textbf{IPW-Random}: The Inverse Propensity Weighting~\cite{DBLP:conf/wsdm/JoachimsSS17, DBLP:conf/sigir/WangBMN16} uses result randomization to estimate the examination probabilities against positions and optimizes the ranking models accordingly. Its performance can be considered as the upper bound for learning-to-rank with user implicit feedback. 
  \item \textbf{Naive}: This model just uses the raw click data to train the ranking model, without any correction. Its performance can be considered as a lower bound for the ranking model.
\end{itemize}

\paragraph{Experimental Protocols}
We implement UPE
and use the baselines in ULTRA~\cite{DBLP:conf/cikm/TranYA21} to conduct our experiments. In particular, UPE is integrated with DLA, as it is the state-of-the-art automatic ULTR algorithm. 
For each query, only the top $N=10$ documents are assumed to be displayed to the users. For both datasets, all models are trained with synthetic clicks. Following the setting in~\cite{DBLP:conf/sigir/AiBLGC18}, the click sessions for training are generated on the fly. We fix the batch size to 256 and train each model for 10K steps. We use the AdaGrad optimizer~\cite{DBLP:journals/jmlr/DuchiHS11} and tune learning rates from 0.01 to 0.05 for each unbiased learning-to-rank algorithm on the validation dataset.

In the experiments, we train neural networks for our ranking functions. All reported results are produced using a model with three hidden layers with size $[512, 256, 128]$ respectively, with the ELU~\cite{DBLP:journals/corr/ClevertUH15} activation function and 0.1 dropout~\cite{DBLP:journals/jmlr/SrivastavaHKSS14}.  
To construct a sufficiently expressive confounder encoder in LPP model, it is configured to have two stacked transformer blocks, each with 256 hidden units and 8 heads. The hidden size of the document and position representation are both set to 64. The feed-forward network is a neural network with two layers with size $[64, 256]$. 

To evaluate all methods, we use the normalized Discounted Cumulative Gain (nDCG)~\cite{DBLP:journals/tois/JarvelinK02} and the Expected Reciprocal Rank (ERR)~\cite{DBLP:conf/cikm/ChapelleMZG09}. For both metrics, we report the results at ranks 1, 3, 5, and 10 to show the performance of models on different positions. Following~\cite{DBLP:conf/sigir/AiBLGC18}, statistical differences are computed based on the Fisher randomization test~\cite{DBLP:conf/cikm/SmuckerAC07} with $p \leq 0.05$.

\begin{figure}
  \centering
  \includegraphics[width=0.95\linewidth]{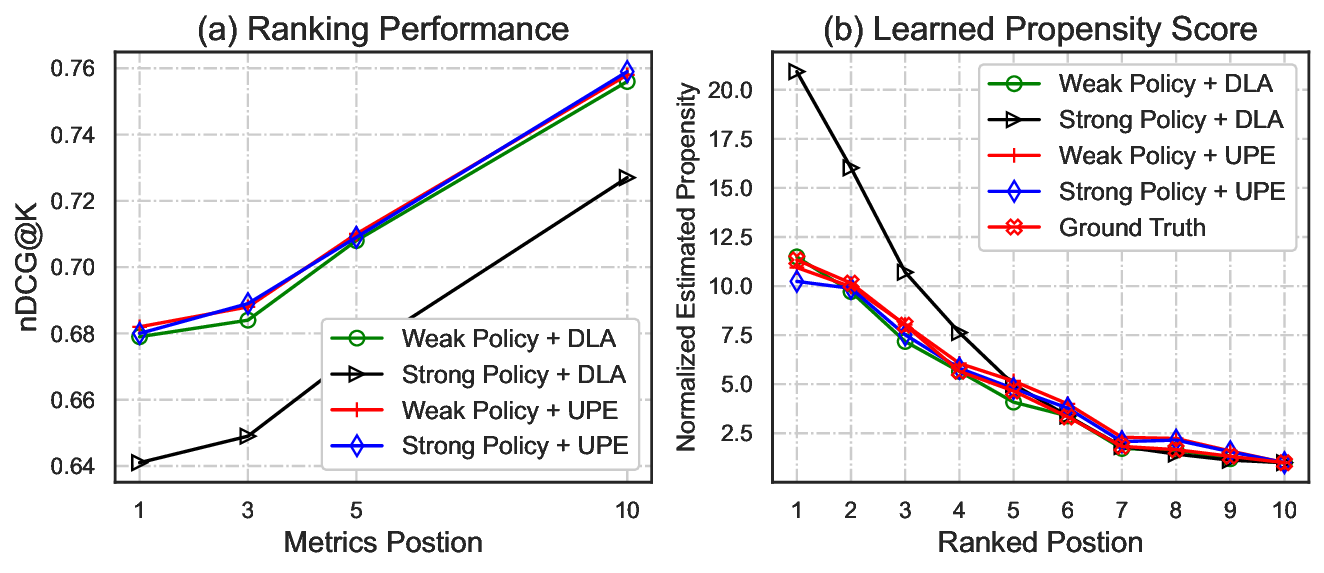}
  \vspace{-10pt}
  \caption{Ranking performance and normalized estimated propensity ($\frac{\text{propensity@K}}{\text{propensity@10}}$) under different logging policies on Yahoo! LETOR. 
  }
  \vspace{-10pt}
  \label{fig:upe_motivation}
\end{figure}
\subsection{Empirical Results for Propensity Overestimation(RQ1)}
\label{sec:upe:motivation}

We first justify the propensity overestimation problem. For a better illustration, we demonstrate the normalized estimated propensities against ranking positions and ranking performance, accordingly.
Fig.~\ref{fig:upe_motivation}(a) investigates the ranking performance of DLA~\cite{DBLP:conf/sigir/AiBLGC18}, a state-of-the-art automatic ULTR method, under weak and strong logging policies, respectively. 
It shows that DLA's ranking performance suffers from a significant drop under a strong logging policy.
Furthermore, in Fig.~\ref{fig:upe_motivation}(b), we show the learned propensity score and indeed observe a propensity overestimation problem in which normalized estimated propensities for top positions are much larger than their actual values. 
This observation confirms the propensity overestimation problem, and this problem is detrimental to ranking models' optimization. Therefore, propensity overestimation needs to be addressed during the dynamic learning and serving of LTR systems, where the logging policies are not longer weak. 

\subsection{Dynamic LTR Simulation (RQ2)}
In practice, the logging policies are updated periodically: we can collect sufficient user clicks in a few months and update the logging policy with the latest ranking policy. 
To simulate this process, we perform deterministic online simulation~($OnD$) and update the logging policy after every 2.5K steps with the most recent ranking model. From Table~\ref{tab:upe:online_exp}, we can see that UPE significantly outperforms all the ULTR baseline methods without result randomization on both datasets. 
This demonstrates the effectiveness of UPE with dynamic training and serving of LTR systems. 
Remarkably, UPE achieves similar performance to IPW-Random, this indicates that UPE can accurately estimate the examination probability for each position, yet without the need for result randomization.

{\centering
\tabcolsep 0.02in
\begin{table*}[!hpt]
\centering
\vspace{-5pt}
\caption{Overall performance comparison between UPE and the baselines on Yahoo! and Istella-S datasets with offline learning~($\mathit{Off}$). ``$\ast$''  indicates statistically significant improvement over the best baseline without result randomization.}
\vspace{-5pt}
\label{tab:upe:offline_exp}
    \centering
    \begin{tabular}{ c c c c c  c c c c c  c c c c c  c c}
    \ChangeRT{0.8pt}
    \multirow{3}{*}{\shortstack{Methods}} & \multicolumn{8}{c}{ Yahoo! LETOR } & \multicolumn{8}{c}{ Istella-S } 
    \\ \cmidrule(lr){2-9} \cmidrule(lr){10-17}
    & \multicolumn{4}{c}{ NDCG@K } & \multicolumn{4}{c}{ ERR@K } & \multicolumn{4}{c}{ NDCG@K } & \multicolumn{4}{c}{ ERR@K }  \\
    \cmidrule(lr){2-5}  
    \cmidrule(lr){6-9}      
    \cmidrule(lr){10-13}  
    \cmidrule(lr){14-17} 
    & K = 1 & K = 3 & K = 5 & K = 10 & K = 1 & K = 3 & K = 5 & K = 10  & K = 1 & K = 3 & K = 5 & K = 10 & K = 1 & K = 3 & K = 5 & K = 10 \\
    \ChangeRT{0.5pt} 
    IPW-Random
          & 0.684 & 0.693 & 0.715 & 0.763 & 0.352 & 0.430 & 0.451 & 0.467
          & 0.676 & 0.640 & 0.664 & 0.723 & 0.604 & 0.710 & 0.727 & 0.733 \\
    \ChangeRT{0.5pt}
    UPE    & $\textbf{0.687}^\ast$ & $\textbf{0.692}^\ast$ & $\textbf{0.715}^\ast$ & $\textbf{0.762}^\ast$
          & $\textbf{0.351}$ & $\textbf{0.429}$ & $\textbf{0.451}$ & $\textbf{0.466}$ 
          & $\textbf{0.669}^\ast$ & $\textbf{0.637}^\ast$ & $\textbf{0.659}^\ast$ & $\textbf{0.720}^\ast$
          & $\textbf{0.598}^\ast$ & $\textbf{0.706}^\ast$ & $\textbf{0.722}^\ast$ & $\textbf{0.729}^\ast$ \\ 
    Vectorization & 0.681 & 0.689 & 0.711 & 0.759 & 0.350 & 0.428 & 0.449 & 0.465 
                  & 0.663 & 0.630 & 0.652 & 0.714 & 0.593 & 0.699 & 0.718 & 0.722  \\
    DLA    & 0.682 & 0.688 & 0.710 & 0.758 & 0.350 & 0.427 & 0.449 & 0.464 
          & 0.662 & 0.629 & 0.653 & 0.713 & 0.592 & 0.699 & 0.717 & 0.723 \\
    REM    & 0.655 & 0.662 & 0.685 & 0.735 & 0.342 & 0.418 & 0.440 & 0.456 
          & 0.609 & 0.570 & 0.589 & 0.637 & 0.545 & 0.653 & 0.672 & 0.681 \\
    PairD  & 0.659 & 0.664 & 0.688 & 0.739 & 0.337 & 0.415 & 0.438 & 0.453 
          & 0.609 & 0.590 & 0.621 & 0.687 & 0.543 & 0.664 & 0.684 & 0.692 \\
    Naive  & 0.662 & 0.666 & 0.688 & 0.739 & 0.338 & 0.416 & 0.438 & 0.454 
          & 0.630 & 0.607 & 0.634 & 0.699 & 0.562 & 0.679 & 0.697 & 0.705 \\ 
    \ChangeRT{0.8pt}
    \end{tabular}
\end{table*}}

\begin{figure}
  \centering
  \includegraphics[width=0.92\linewidth]{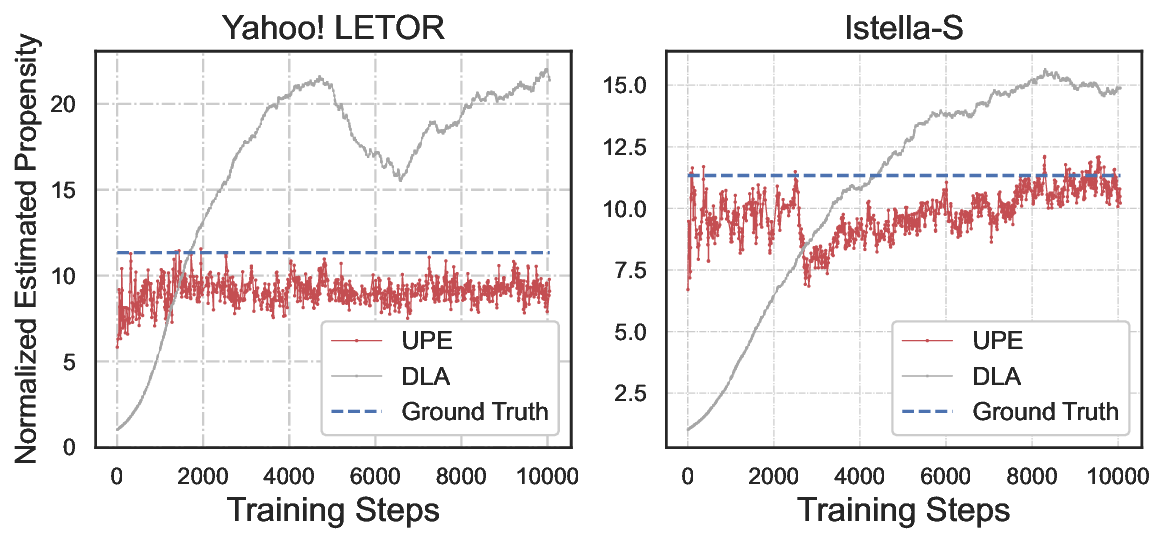}
  \vspace{-10pt}
  \caption{The learning curve of normalized estimated propensity for position 1 with deterministic online learning ($\mathit{OnD}$). 
  }
  \vspace{-10pt}
  \label{fig:upe:online_propensity_curve}
\end{figure}

To provide a more insightful understanding for the benefit of UPE, we also illustrate the learning curve of estimated propensity in Fig.~\ref{fig:upe:online_propensity_curve}.  In particular, we investigate the normalized estimated propensity against position 1 on UPE, compared with DLA, the state-of-the-art automatic ULTR algorithm.  We select position 1 for illustration because it suffers from propensity overestimation most (will see Fig.~\ref{fig:upe_motivation}).
The ``ground truth'' is computed by $\frac{\rho_1}{\rho_{10}}$, where $\rho_i$ is the position bias defined in Eq.\ref{eq:upe:rho}. 

We can see that DLA suffers from propensity overestimation, as evidenced by its learning curves deviating from the ground truth during dynamic training. This demonstrates that existing ULTR methods are unable to leverage implicit feedback to enhance the ranking model during dynamic training and serving of LTR systems, due to the issue of propensity overestimation. 
In contrast, UPE's propensity estimation is robust to changes in the logging policy, and its curves generally align with the ground truth. As more user clicks are collected, the ranking performance improves, highlighting the effectiveness of UPE in enhancing the ranking model during training.
This nice property actualizes the value of implicit feedback in the practical ULTR scenario, \ie the LTR systems are updated dynamically.
%
Remarkably, though the performance of logging policy varies from scratch to being in high accuracy, UPE can accurately estimate propensity in a consistent manner. 
This suggests that the superiority of UPE is not related to the performance of logging policies, but the accurate estimation for confounding effect under a logging policy, which justifies the necessity of an expressive confounder encoder in LPP. 

It is worth explaining a counter-intuitive observation: the proposed UPE outperforms the ideal IPW-Random~(theoretical upper bound), which uses result randomization to estimate the examination probability against positions. While result randomization would produce the accurate estimation of the true position bias and leads to the optimality of IPW-Random in theory, it also introduces large variance in practice. The ``unexpected'' results observed in Table~\ref{tab:upe:online_exp} and Table~\ref{tab:upe:offline_exp} are mostly due to such variances.

\subsection{Offline Simulation (RQ3)}
\label{section:upe:offline}

To investigate the generalizability of UPE, we conduct experiments on the commonly used ULTR setting, \ie offline learning paradigm~($\mathit{Off}$) as in~\citet{DBLP:conf/wsdm/JoachimsSS17} and~\citet{DBLP:conf/sigir/AiBLGC18}. Unlike the online paradigm, the ranked lists are generated by a Rank SVM model~\cite{DBLP:conf/kdd/Joachims06} that trained with $1\%$ of the training data with real relvance judgements, \ie weak logging policy.

Table~\ref{tab:upe:offline_exp} shows the experimental results. We can see that UPE also achieves the best performance among all baseline methods without result randomization, and UPE achieves similar performance as IPW-Random on the offline paradigm $\mathit{Off}$. 
Besides, UPE outperforms the best baseline methods without result randomization in most of the metrics. This observation indicates that UPE can effectively address propensity overestimation even when there exists a minor confounding effect by query-document relevance confounder $X$ under a weak logging policy, and more importantly such confounding effect is indeed non-negligible.

To validate the effectiveness of UPE in propensity estimation under different logging policies, we have also demonstrated the overall distribution of normalized estimated propensity against position in Fig.~\ref{fig:upe_motivation}. The results demonstrate the high generalizability of UPE, which can consistently obtain unconfounded propensity estimation under both strong and weak policies. 
Its ranking performance is also shown in Fig.~\ref{fig:upe_motivation}, where UPE can consitenly achieve the SOTA ranking performance under both weak and strong policies.



\subsection{Ablation Study (RQ4)}
Our ablation studies investigate how variant designs of LPP affect its optimization, and consequentially affect the ranking performance.

\subsubsection{Is it more beneficial to model $P(\hat{R} | \mathbf{x})$ than $P(K|\mathbf{x})$ in propensity model training?}
\label{section:upe:target}
In this section, we compare different fitting targets for LPP optimization. To instantiate $P(K|\mathbf{x})$, we transform the ranked positions as relevance labels for ranking optimization as in~\citet{DBLP:conf/cikm/ZhangMLZ0MXT19}:
\begin{subequations}
    \begin{align}
    \textbf{MRR-UPE} &: \ \ \text{MRR-LPP}@K = \frac{1}{K}  \\
    \textbf{DCG-UPE} &: \ \ \text{DCG-LPP}@K = \frac{1}{\log_2{(K+1)}}.
    \end{align}
    \vspace{-.1in}
\end{subequations}
{\centering
\tabcolsep 0.02in
\begin{table}[!pt]
\centering
\caption{The performance of different fitting targets in LLP optimization on Yahoo! LETOR with deterministic online learning~($\mathit{OnD}$). $``\ast"$  indicates statistically significant improvement over the best baseline.}
\vspace{-5pt}
\label{tab:upe:LLP_variant}
    \centering
    \small{
    \begin{tabular}{ c c c c c  c c c c}
    \ChangeRT{0.8pt}
    \multirow{2}{*}{\shortstack{Methods}} & \multicolumn{4}{c}{ NDCG@K } & \multicolumn{4}{c}{ ERR@K }  \\
    \cmidrule(lr){2-5}  
    \cmidrule(lr){6-9}      
    & K = 1 & K = 3 & K = 5 & K = 10 & K = 1 & K = 3 & K = 5 & K = 10 \\
    \ChangeRT{0.5pt} 
    UPE    & $\textbf{0.695}^\ast$ & $\textbf{0.703}^\ast$ & $\textbf{0.723}^\ast$ & $\textbf{0.768}^\ast$
           & $\textbf{0.355}^\ast$ & $\textbf{0.433}^\ast$ & $\textbf{0.454}^\ast$ & $\textbf{0.469}^\ast$ \\ 
    MRR-UPE & 0.688 & 0.694 & 0.715 & 0.762 & 0.352 & 0.429 & 0.451 & 0.466 \\
    DCG-UPE & 0.686 & 0.691 & 0.714 & 0.760 & 0.350 & 0.428 & 0.451 & 0.465 \\
    \ChangeRT{0.8pt}
    \end{tabular}}
    \vspace{-10pt}
\end{table}}

The experimental results are summarized in Table~\ref{tab:upe:LLP_variant}. We can see that UPE significantly outperforms the two variants, \ie MRR-UPE and DCG-UPE. This observation confirms that the logging policy scores are more informative than the ranked positions, because logging policy scores not only provide the order of the documents, \ie ranked positions, but also their relevance strengths.

\subsubsection{Is the two-step optimization strategy, \ie logging-policy-aware confounding effect learning and joint propensity learning, indispensable for LPP optimization?} 
\label{section:upe:wo_tst}
We argue that existing automatic ULTR methods cannot extract separated ranking model $P(R|X)$ and propensity model $P(E|K, X)$ from raw click data in UPE. In this section, we justify this argument, and demonstrate the indispensability of our novel two-step optimization. 

\begin{figure}
  \centering
  \includegraphics[width=0.99\linewidth]{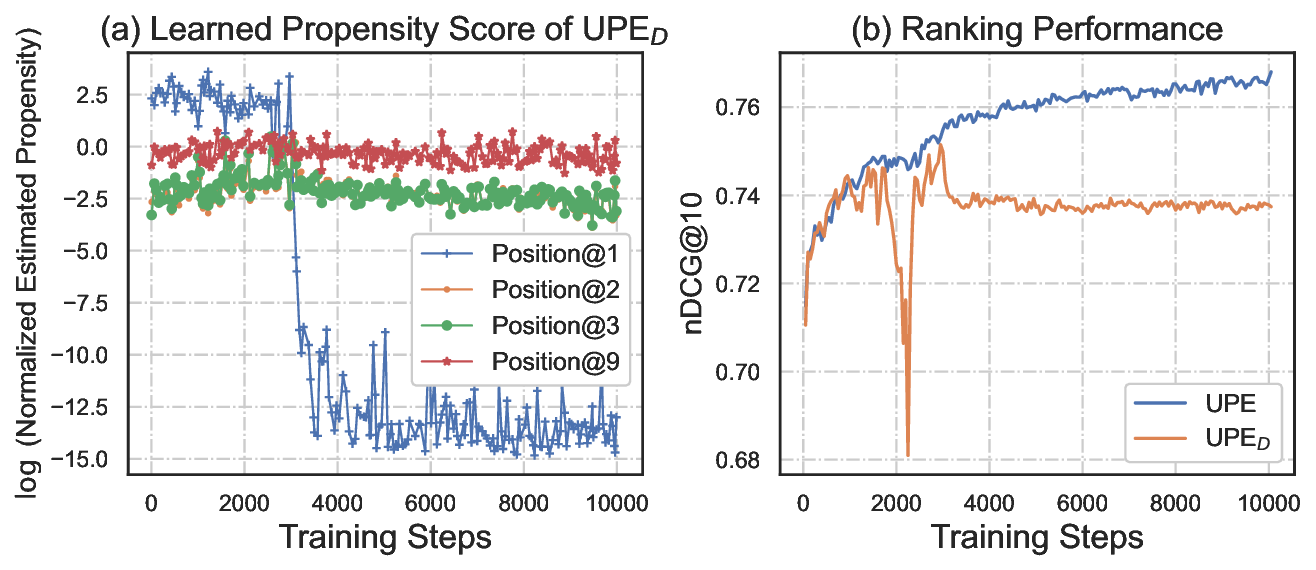}
  \vspace{-10pt}
  \caption{The learning curve of normalized estimated propensity and ranking performance of UPE and UPE$_D$ on Yahoo! LETOR with deterministic online learning ($\mathit{OnD}$).
  }
  \vspace{-10pt}
  \label{fig:upe:two_step}
\end{figure}

We refer to the naive propensity learning framework as UPE$_N$, which optimizes $P(R|X)$ and $P(E|K, X)$ from raw click data alternatively. For better illustration, we show the logarithm of the normalized estimated propensity at positions 1, 2, 3 and 9 in Fig.~\ref{fig:upe:two_step}(a).  
We can observed that from step $3000$, the estimated propensity by UPE$_N$ against position 1 has been much smaller than that against position 9, and those at other positions are almost identical to that at position 9. It indicates that UPE$_N$ fails to learn the correct examination probability for each position, which should have a larger value for a higher-ranked position.
We side-by-side present the ranking performance curve of UPE$_N$ in Fig.~\ref{fig:upe:two_step}(b). Starting from step $3000$, nDCG@10 of UPE$_N$ does not increase; this means that collecting more click data does not improve the ranking model because the propensity model has failed in separating the impact of relevance and position on
propensity. This phenomenon also verifies that an accurate propensity model is indeed necessary for optimizing ranking models.

Unlike UPE$_D$, the proposed logging-policy-aware confounding effect learning and joint propensity learning enable us to obtain an unconfounded propensity through backdoor adjustment, which correctly estimates the causal effect between the propensity and position, as has been shown in Fig.~\ref{fig:upe:online_propensity_curve}. 
Moreover, the learning curve of UPE in Fig.~\ref{fig:upe:two_step}(b) shows that the ranking performance of our proposal consistently improves when more click data are collected.
Therefore, the two-step optimization strategy is necessary for LPP optimization, it enables the separation of ranking and propensity models from raw clicks.


\section{Conclusion}

In this work, we investigate unbiased learning to rank through the lens of causality and identify query-document representation as a confounder, which leads to propensity overestimation.
For unconfounded propensity overestimation, we propose a novel propensity model, \ie Logging-Policy-aware Propensity Model, and its distinct two-step optimization strategy: 
(1) logging-policy-aware confounding effect learning, which captures the confounding effect caused by the query-document feature confounder and thereby separates ranking and propensity models from raw clicks;
and (2) joint propensity learning, which learns the mapping from position and query-document feature to the examination by fixing the confounding effect model and solely tuning the position-related parameters. 
Given the fine-tuned LPP model, we conduct backdoor adjustment for unconfounded propensity estimation, which serves for ULTR.
Extensive experiments on two benchmarks with synthetic clicks with online and offline simulations validate the effectiveness of our proposal in addressing propensity overestimation and improving ranking performance. 
A natural future direction would be to extend current work to pairwise learning to explore the feasibility of UPE across more ULTR frameworks.

\bibliographystyle{ACM-Reference-Format}\balance
\bibliography{sample-base}

\end{document}